\documentclass[prl,amsmath,amssymb,amsfonts,twocolumn]{revtex4}
\usepackage{amsfonts}
\usepackage{amsmath}

\usepackage[pdftex]{graphicx}

\begin{document}

\title{Optical Quantum Computing}
\author{Jeremy L. O'Brien}
\email{Jeremy.OBrien@bristol.ac.uk}
\affiliation{Centre for Quantum Photonics, H. H. Wills Physics Laboratory \& Department of Electrical and Electronic Engineering, University of Bristol, Merchant Venturers Building, Woodland Road, Bristol, BS8 1UB, UK}

\begin{abstract}
In 2001 all-optical quantum computing became feasible with the discovery that scalable quantum computing is possible using only single photon sources, linear optical elements, and single photon detectors. Although it was in principle scalable, the massive resource overhead made the scheme practically daunting. However, several simplifications were followed by proof-of-principle demonstrations, and recent approaches based on cluster states or error encoding have dramatically reduced this worrying resource overhead, making an all-optical architecture a serious contender for the ultimate goal of a large-scale quantum computer. Key challenges will be the realization of high-efficiency sources of indistinguishable single photons, low-loss, scalable optical circuits, high efficiency single photon detectors, and low-loss interfacing of these components.
\end{abstract}

\maketitle 

Over the last few decades quantum information science has emerged to consider what additional power and functionality can be realised in the encoding, transmission and processing of information by specifically harnessing quantum mechanical effects \cite{nielsen}. Anticipated technologies include: quantum key distribution \cite{gi-rmp-74-145}, which offers perfectly secure communication; quantum metrology \cite{gi-sci-306-1330}, which allows more precise measurements than could ever be achieved without quantum mechanics; and quantum litography \cite{bo-prl-85-2733}, which could enable fabrication of devices with features much smaller than the wavelength of light. Perhaps the most startling and powerful future quantum technology is a quantum computer, which promises exponentially faster computation for particular tasks \cite{nielsen,de-prsla-400-97}.

The quest to develop a quantum computer will require formidable technical mastery of the fabrication of devices at the nano and possibly atomic scale, and precision control of their quantum mechanical states. The task is also daunting owing to the inherent fragility of quantum states and the fact that quantum entanglement, and its role in a quantum computer, is not yet fully understood. As we engineer devices that exploit quantum mechanical effects, we will gain an unprecedented control over the fundamental workings of nature as well as a deeper understanding of them.

The requirements for realizing a quantum computer are confounding: scalable physical qubits---two state quantum systems---that can be well isolated from the environment, but also initialised, measured, and controllably interacted to implement a universal set of quantum logic gates \cite{di-sm-23-419}. However, a number of physical implementations are being pursued, including nuclear magnetic resonance, ion, atom, cavity quantum electrodynamics, solid state, and superconducting  systems \cite{qcroadmap}. Over the last few years single particles of light---photons---have emerged as one of several leading approaches \cite{qcroadmap}.

\noindent\textbf{Single Photons as Qubits} 

\begin{figure}
\begin{center}
\includegraphics*[width=0.5\textwidth]{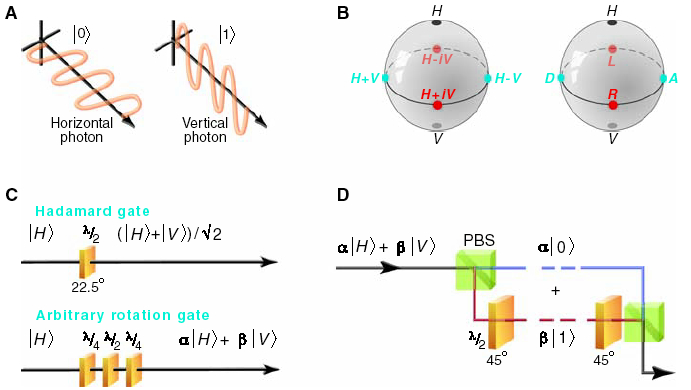}
\caption{\noindent {\bf Fig. 1.} Single photon qubits. ({\bf A}) A horizontal (H) photon represents a logical ``0" and a vertical (V) photon represents a logical ``1": $|0\rangle\equiv|H\rangle$; $|1\rangle\equiv|V\rangle$. ({\bf B}) An arbitrary state can be plotted on the Bloch (or Poincar\'{e}) sphere. Examples of diagonal ($|D\rangle\equiv|0\rangle+|1\rangle$), anti-diagonal ($|A\rangle\equiv|0\rangle-|1\rangle$), right circular ($|R\rangle\equiv|0\rangle+i|1\rangle$), and left circular ($|L\rangle\equiv|0\rangle-i|1\rangle$) are shown. ({\bf C}) Single qubit gates are easily realized using birefringent waveplates that retard one polarization by a fraction of a wavelength $\lambda$ relative to an orthogonal polarization, causing a rotation of the state on the Bloch sphere, with the axis of rotation determined by the orientation of the waveplate. For example, a Hadamard (H) gate (defined by its operation on the logical states: $|0\rangle\rightarrow|0\rangle+|1\rangle$; $|1\rangle\rightarrow|0\rangle-|1\rangle$ or $|H\rangle\rightarrow|D\rangle$; $|V\rangle\rightarrow|A\rangle$) causes a $\pi$ rotation about an axis running through the midpoint between the $|H\rangle$ and $|D\rangle$ states, and can be realized by a $\lambda/2$ waveplate oriented at 22.5$^{\circ}$. An arbitrary rotation requires a $\lambda/4$--$\lambda/2$--$\lambda/4$ sequence. ({\bf D}) Converting between polarization and path encoding requires a polarizing beam splitter (PBS), which transmits $H$ and reflects $V$, and a $\lambda/2$ waveplate oriented at 45$^{\circ}$, which transforms $|V\rangle\leftrightarrow|H\rangle$.
}
\label{fig1}
\end{center}
\end{figure}

\noindent Single photons are largely free of the noise---or decoherence---that plagues other systems; can be easily manipulated to realize one-qubit logic gates; and enable encoding in any of several degrees of freedom---polarization, time bin, path, \emph{etc}.  Figure 1A shows how a qubit can be encoded in the polarization of a single photon. An arbitrary state of a single qubit $\alpha|H\rangle+\beta|V\rangle$ ($|\alpha|^2+|\beta|^2=1$) can be represented on the Poincar\'{e} (or Bloch) sphere (Fig. 1B). One-qubit logic gates are straightforward using birefringent waveplates (Fig. 1C) and converting between polarisation and path encoding can be easily achieved using a polarizing beam splitter (Fig. 1 D), where $|0\rangle$ or $|1\rangle$ now represents a photon in the upper or lower path, respectively. 
  
A major difficulty for optical quantum computing is in realizing the entangling logic gates required for universal quantum computation. The canonical example is the controlled-NOT gate (CNOT), which flips the state of a target (T) qubit conditional on a control (C) qubit being in the logical state ``1". Figure 2A shows why this operation is difficult.  The two paths used to encode the target qubit are mixed at a 50\% reflecting beamsplitter (BS) (or half-silvered mirror), which performs the Hadamard operation (see Fig. 1C caption). If the phase shift is not applied, the second Hadamard (BS) undoes the first, returning the target qubit to exactly the same state it started in (this is an example of (classical) wave interference). If, however, a ($\pi$) phase shift is applied---\emph{i.e.} $|0\rangle+|1\rangle\leftrightarrow|0\rangle-|1\rangle$---the target qubit undergoes a bit-flip or NOT operation. A CNOT must implement this phase shift only if the control photon is in the ``1" path. No known or foreseen material has an optical non-linearity strong enough to implement this conditional phase shift (although tremendous progress has been made with single atoms in high-finesse optical cavities \cite{tu-prl-75-4710,ao-nat-443-671,hi-nphys-3-253}). 

In 2001 a major breakthrough showed that scalable quantum computing is possible using only single photon sources and detectors, and simple (linear) optical circuits consisting of BSs \cite{kn-nat-409-46}. This is a truly remarkable discovery since the argument above suggests that a strong optical non-linearity is required to realize the most basic logic element.

\begin{figure}[t]
\begin{center}
\includegraphics*[width=0.5\textwidth]{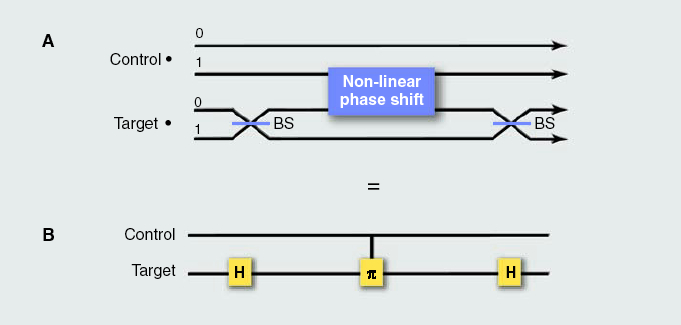}
\caption{\noindent {\bf Fig. 2.} An optical controlled-NOT gate. ({\bf A}) Schematic of a possible realization of an optical CNOT gate. See text for details. ({\bf B}) In the notation of quantum circuits the BSs implement a Hadamard (H) gate.
}
\label{fig2}
\end{center}
\end{figure}

\noindent\textbf{Linear Optical Quantum Computing} 

\begin{figure}[t]
\begin{center}
\includegraphics*[width=0.5\textwidth]{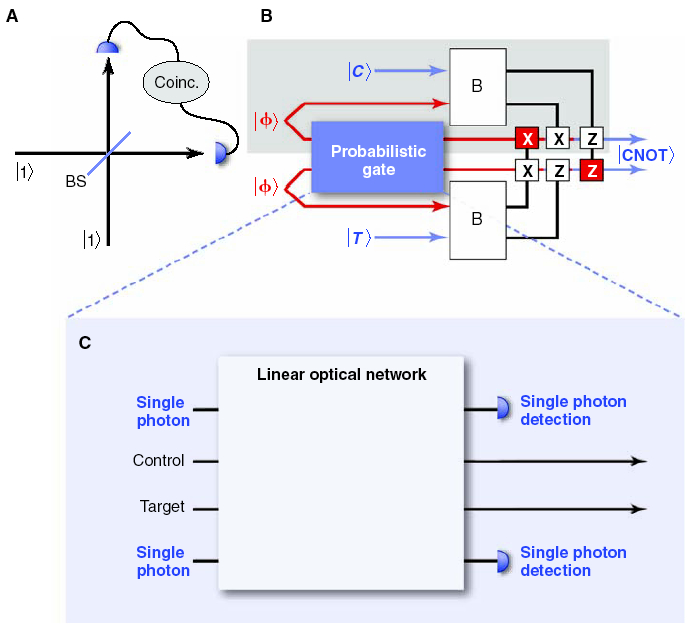}
\caption{\noindent {\bf Fig. 3.} An optical CNOT gate via teleportation (\textbf{A}) A cartoon of a measurement assisted nondeterministic CNOT gate. (\textbf{B}) Quantum interference of two photons at a BS. (\textbf{C}) Teleportation of a CNOT: Ignoring the ``Probabilistic Gate", a qubit in an unknown state $|C\rangle$ and one of two photons prepared in a maximally entangled state $|\phi\rangle$ are subjected to a Bell measurement (B). This measurement leaves the third qubit in the state $|C\rangle$ or the bit (X) and/or phase (Z) flipped version of $|C\rangle$, depending on which of the four maximally entangled states is measured. An unwanted X and/or Z flip can be trivially corrected by applying a second X and/or Z as required. Still ignoring the ``Probabilistic Gate", the unknown input state of the control and target qubits can both be teleported and the CNOT performed on the output qubits. This seems like a lot of extra work for no gain, however, performing the CNOT before the (possible) X and Z flip has the tremendous advantage that we could repeatedly attempt the CNOT on the two halves of two entangled states, and only when the gate works would we proceed with teleportation. In this way the control and target qubits are preserved until the gate works (on average 32 entangled states will be consumed) and we can implement the CNOT deterministically.  In quantum mechanics the order in which operations are performed is important; performing the CNOT earlier means we must add the X and Z flips indicated in red in Fig. 3C  \cite{go-nat-402-390}. 
}
\label{fig3}
\end{center}
\end{figure}

\noindent A cartoon of a non-deterministic (probabilistic with success signal)  CNOT is shown in Fig. 3A. The control and target qubits (encoded in polarization, say), together with two auxiliary photons, enter an optical network of BSs, where the four photons' paths are combined. At the output of this network, the control and target photons emerge, having had the CNOT logic operation applied to their state, conditional on a single photon being detected at both detectors. This detection event occurs with probability $P<1$ (1/16 in the original scheme); the rest of the time ($P=15/16$) another detection pattern is recorded (none, only one, two photons at one detector, etc) and the CNOT logic is not applied---in fact a single photon may not even emerge from the control and/or target outputs in these cases. 

A nondeterministic CNOT  is of little use for quantum computing as the probability that a computation succeeds decreases exponentially with the number of CNOTs. Fortunately, the success probability of the nondeterministic CNOT can be boosted by harnessing quantum teleportation \cite{be-prl-70-1895}---a process whereby the unknown state of a qubit can be transfered to another qubit. The idea is to teleport a non-deterministic gate that has already worked onto the control and target qubits \cite{go-nat-402-390} (see Fig. 3 caption for details). Quantum teleportation has been realized with single photons \cite{bo-nat-390-575}.

An important omission from the above discussion is that since the Bell measurements required for teleportation (Fig. 3C) measure maximally entangled states they requires a similar optical nonlinearity to a CNOT (although the photons can be destroyed in the measurement) and therefore fail some of the time. When they fail, they measure the state of the control and target photons in the $\{|0\rangle,|1\rangle\}$ basis. The final component is an encoding against this ``measurement error": a single logical qubit is encoded in several physical qubits such that if one of the physical qubits is measured, the original logical qubit can still be recovered. These encoded states are entangled and therefore require entangling gates to realize them. However, by using more and more photons a CNOT with success probability approaching one can be realized \cite{kn-nat-409-46}.

\noindent\textbf{Reducing the Resource Overhead} 

\noindent These developments were expanded upon \cite{ko-pra-63-030301,pi-pra-64-062311,ra-pra-65-012314,ra-pra-65-062324,ho-pra-66-024308} and soon followed by several proof-of-principal experimental demonstrations of CNOTs \cite{pi-pra-68-032316,ob-nat-426-264, ob-prl-93-080502, ga-prl-93-020504} and encoding against measurement error \cite{ob-pra-71-060303,pi-pra-71-052332}. Despite this great progress, optical quantum computing was still widely regarded as impractical owing to the large resource overhead required to realize a near-deterministic CNOT: $>$10,000 pairs of entangled photons to achieve a success probability of $>$95\%. The reason that all-optical quantum computing is today a promising route to practical quantum computing is due to new schemes that dramatically reduce this worrying resource overhead.

Quantum computations (regardless of physical realization) are typically formulated using the quantum circuit model (eg. Fig. 2B),  a generalization of the circuit model for Boolean logic: qubits are represented by wires propagating in time from left to right, subjected to a sequence of quantum logic gates, and finally measured \cite{nielsen}. In 2001 a remarkable alternative was proposed in which the computation starts with a particular massively entangled state of many qubits---a cluster state---and the computation proceeds via a sequence of single qubit measurements from left to right that ultimately leave the rightmost column of qubits in the answer state \cite{ra-prl-86-5188} (Fig. 4).

In 2004 it was recognized that the cluster approach offered tremendous advantages for optical realizations \cite{ni-prl-93-040503} (see also \cite{yo-prl-91-037903}). Because preparation the cluster state can be probabilistic, non-deterministic CNOTs are suitable for making it, removing much of the massive overhead that arises from the error encoding used to make near-deterministic CNOTs. It turns out that a similar advantage can also be gained in the circuit forumulation of optical quantum computing by using more sophisticated error encoding techniques \cite{ra-prl-95-100501}. These, and other techniques that dispense with CNOT gates entirely \cite{br-prl-95-010501}, reduce the resources required 
by 3-4 orders of magnitude, making an all-optical approach far more attractive. There have already been experimental proof-of-principle demonstrations of these new schemes (\emph{eg.} \cite{wa-nat-434-169,ki-prl-95-210502,pr-nat-445-65,lu-nphys-3-91}).

\begin{figure}[t]

\begin{center}
\includegraphics*[width=0.48\textwidth]{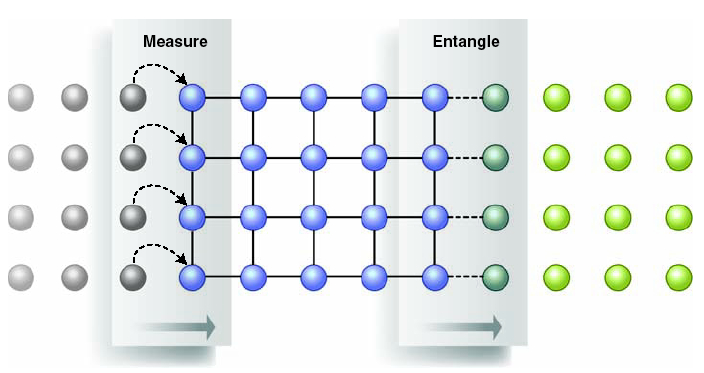}
\caption{\noindent {\bf Fig. 4.} Cluster state quantum computing. For photons it is practical to start measuring the qubits, while the cluster is still being grown: The blue qubits are in a cluster state, where the bonds between them represent entanglement. The green qubits are being added to the cluster, while the grey qubits have been measured and are no longer entangled. The measurement outcome determines the basis for the measurement on the next qubit.
}
\label{fig4}
\end{center}
\end{figure}

\noindent\textbf{Fault Tolerance} 

\noindent The final, and arguably most important consideration (for all physical realizations), is fault tolerance \cite{nielsen}. In contrast to conventional computers, quantum computers will be very susceptible to noise, which must be encoded against (in addition to the encoding described above). The threshold theorem says that if the noise is below some threshold an arbitrarily long quantum computation can be realized. One of the most encouraging results for all approaches to quantum computing was the high threshold of 1\% recently reported by Knill \cite{kn-nat-434-39}. Because cluster state approaches do not conform to the standard model the threshold theorem does not apply; fortunately analogous thresholds have been shown to exist \cite{ni-pra-71-042323}. Recent results give cause for optimism: They show that if the product of source and detector efficiency is $>2/3$ then optical quantum computing is possible, provided all other components operate perfectly \cite{va-quantph-0702044} (photon loss can in some cases be incorporated into source or detector efficiency). More complete treatments that consider more sources of noise give thresholds of $10^{-3}-10^{-4}$ \cite{da-pra-73-052306}. The true number will likely lie somewhere in between.

\noindent\textbf{Sources, Detectors, and Circuits} 

\noindent There are very stringent requirements for single photon sources for optical quantum computing. In a general linear optical network (\textit{eg.} Fig. 3A) there are places where photons arrive at both inputs to a BS where quantum interference of two (or more) photons can occur. An example is shown in Fig. 3B where a photon enters each input of a 50\% reflective BS.  
The probability of detecting a single photon at each output is given by the square of the sum of the probability amplitude for both photons to be transmitted and that for both photons to be reflected: $P=|r.r+t.t|^2$. Because a phase shift occurs on reflection $r.r=-t.t$ and so $P=0$, in contrast to our (classical) expectation: $P=1/2$  \cite{scirevnote}.  
In order for quantum interference to occur the two photons must be indistinguishable from one another in all degrees of freedom. 

To date, small scale tests of optical quantum computing have relied on indistinguishable pairs of photons generated by a strong laser pulse in a non-linear crystal. Unfortunately this process is spontaneous and not readily scalable \cite{sps}. Solid state sources of single photons hold the promise of ready integration, and quantum interference between subsequent photons emitted from a semiconductor quantum dot has been observed \cite{sa-nat-419-594}. However, an optical quantum computer will require quantum interference between photons emitted from independent sources. This has very recently been achieved for a pair of trapped atoms \cite{be-nat-440-779} and ions \cite{ma-nphys-3-538}, a tremendous advance that bodes well for optical quantum computing. Impurities in diamond may offer the best of both worlds---a solid state host and atom-like energy levels---and have emerged as very promising candidates \cite{sps}.
 
It is actually the inherent non-linarity of photon measurement, combined with quantum interference of photons, that makes linear optical quantum computing possible.  Single photon counting modules are commercially available and have been used almost all demonstrations to date, however, they cannot distinguish between one or more photons and have a limited efficiency ($\sim$70\%). Higher efficiency will be required for scalable optical quantum computing, while photon number resolution will be desirable. Ongoing work indicates that such high performance detectors will become available, with superconductor-based devices holding great promise \cite{spd}.

Finally, almost all demonstrations of linear optical logic circuits have relied on large scale BSs and mirrors, with photons propagating in air;  improved performance,  miniaturisation, and scalability will likely require low-loss microscopic optical waveguide circuits. A promising approach is integrated optics---an analogue of electrical integrated circuits---which has been developed by the photonics industry. Outstanding challenges are to realize quantum interference in these devices and to integrate them with single photon sources and detectors.

\noindent\textbf{Nonlinear and Hybrid Approaches} 

\noindent Recently attention has been given to the idea of combining linear optics with optical nonlinearities that would not allow a CNOT gate to be realized in the fashion suggested in Fig. 2, but would nevertheless offer significant advantages. One is to use a two-photon absorber to implement the quantum Zeno effect, whereby repeated measurement inhibits the emission of two photons into one of the outputs of a CNOT gate \cite{fr-pra-70-062302}---the failure mode of the linear optical CNOT gate proposed in \cite{pi-pra-64-062311}. Another is to use a strong optical nonlinearity that is significantly weaker than that required in Fig. 2: single photons are made to interact with one-another via a bright laser pulse and the nonlinear medium \cite{ba-pra-71-060302}. Finally, recent developments suggest that a hybrid approach may have many advantages \cite{li-pra-73-012304}: because single photon sources are inherently quantum mechanical, it is promising to consider storing quantum information in the sources themselves; already spins associated with impurities in diamond have shown great promise in this direction \cite{sps,mbqc}. Such systems are particularly suited to the small-scale quantum processors that will be required in the nodes and quantum repeaters of quantum communication networks.

\noindent\textbf{Future Prospects} 

\noindent Despite great progress, much work remains to be done if a large-scale optical quantum computer is to be realized. It is not yet known whether the circuit or cluster model (or some other approach) is most promising; indeed a combination of these approaches where error encoding is achieved using cluster techniques but the computation proceeds via conventional CNOT gates has been described \cite{gi-pra-75-052328}. Further, the role of nonlinear optics approaches in any future optical quantum computer will depend on their efficacy and practicality.  The majority of experimental demonstrations to date have relied on a non-scalable single photon sources, large-scale optical elements, and modest efficiency single photon detectors;  scaling to useful devices will require high efficiency single photon sources and detectors that are efficiently coupled to low-loss microscopic optical waveguide circuits (optical memories may not be required \cite{ki-quantph-0611140}).


\begin{thebibliography}{10}

\bibitem{nielsen}
M.~A. Nielsen, I.~L. Chuang, {\it Quantum Computation and Quantum
  Information\/} (Cambridge University Press, 2000).

\bibitem{gi-rmp-74-145}
N.~Gisin, G.~Ribordy, W.~Tittel, H.~Zbinden, {\it Rev. Mod. Phys.\/} {\bf 74},
  145 (2002).

\bibitem{gi-sci-306-1330}
V.~Giovannetti, S.~Lloyd, L.~Maccone, {\it Science\/} {\bf 306}, 1330 (2004).

\bibitem{bo-prl-85-2733}
A.~N. Boto, {\it et~al.\/}, {\it Phys. Rev. Lett.\/} {\bf 85}, 2733 (2000).

\bibitem{de-prsla-400-97}
D.~Deutsch, {\it Proc. R. Soc. Lond. A\/} {\bf 400}, 97 (1985).

\bibitem{di-sm-23-419}
D.~P. DiVincenzo, D.~Loss, {\it Superlatt. Micro.\/} {\bf 23}, 419 (1998).

\bibitem{qcroadmap}
See the US Advanced Research and Development Activity (ARDA) Quantum
  Computation Roadmap for current state-of-the-art:
  $http://qist.lanl.gov/qcomp_map.shtml$.

\bibitem{tu-prl-75-4710}
Q.~A. Turchette, C.~J. Hood, W.~Lange, H.~Mabuchi, H.~J. Kimble, {\it Phys.
  Rev. Lett.\/} {\bf 75}, 4710 (1995).

\bibitem{ao-nat-443-671}
T.~Aoki, {\it et~al.\/}, {\it Nature\/} {\bf 443}, 671 (2006).

\bibitem{hi-nphys-3-253}
M.~Hijlkema, {\it et~al.\/}, {\it Nature Phys.\/} {\bf 3}, 253 (2007).

\bibitem{kn-nat-409-46}
E.~Knill, R.~Laflamme, G.~J. Milburn, {\it Nature\/} {\bf 409}, 46 (2001).

\bibitem{be-prl-70-1895}
C.~H. Bennett, {\it et~al.\/}, {\it Phys. Rev. Lett.\/} {\bf 70}, 1895 (1993).

\bibitem{go-nat-402-390}
D.~Gottesman, I.~L. Chuang, {\it Nature\/} {\bf 402}, 390 (1999).

\bibitem{bo-nat-390-575}
D.~Bouwmeester, {\it et~al.\/}, {\it Nature\/} {\bf 390}, 575 (1997).

\bibitem{ko-pra-63-030301}
M.~Koashi, T.~Yamamoto, N.~Imoto, {\it Phys. Rev. A\/} {\bf 63}, 030301 (2001).

\bibitem{pi-pra-64-062311}
T.~B. Pittman, B.~C. Jacobs, J.~D. Franson, {\it Phys. Rev. A\/} {\bf 64},
  062311 (2001).

\bibitem{ra-pra-65-012314}
T.~C. Ralph, A.~G. White, W.~J. Munro, G.~J. Milburn, {\it Phys. Rev. A\/} {\bf
  65}, 012314 (2001).

\bibitem{ra-pra-65-062324}
T.~C. Ralph, N.~K. Langford, T.~B. Bell, A.~G. White, {\it Phys. Rev. A\/} {\bf
  65}, 062324 (2001).

\bibitem{ho-pra-66-024308}
H.~F. Hofmann, S.~Takeuchi, {\it Phys. Rev. A\/} {\bf 66}, 024308 (2001).

\bibitem{pi-pra-68-032316}
T.~B. Pittman, M.~J. Fitch, B.~C. Jacobs, J.~D. Franson, {\it Phys. Rev. A\/}
  {\bf 68}, 032316 (2003).

\bibitem{ob-nat-426-264}
J.~L. O'Brien, G.~J. Pryde, A.~G. White, T.~C. Ralph, D.~Branning, {\it
  Nature\/} {\bf 426}, 264 (2003).

\bibitem{ob-prl-93-080502}
J.~L. O'Brien, {\it et~al.\/}, {\it Phys. Rev. Lett.\/} {\bf 93}, 080502
  (2004).

\bibitem{ga-prl-93-020504}
S.~Gasparoni, J.-W. Pan, P.~Walther, T.~Rudolph, A.~Zeilinger, {\it Phys. Rev.
  Lett.\/} {\bf 93}, 020504 (2004).

\bibitem{ob-pra-71-060303}
J.~L. O'Brien, G.~J. Pryde, A.~G. White, T.~C. Ralph, {\it Phys. Rev. A\/} {\bf
  71}, 060303 (2005).

\bibitem{pi-pra-71-052332}
T.~B. Pittman, B.~C. Jacobs, J.~D. Franson, {\it Phys. Rev. A\/} {\bf 71},
  052332 (2005).

\bibitem{ra-prl-86-5188}
R.~Raussendorf, H.~J. Briegel, {\it Phys. Rev. Lett.\/} {\bf 86}, 5188 (2001).

\bibitem{ni-prl-93-040503}
M.~A. Nielsen, {\it Phys. Rev. Lett.\/} {\bf 93}, 040503 (2004).

\bibitem{yo-prl-91-037903}
N.~Yoran, B.~Reznik, {\it Phys. Rev. Lett.\/} {\bf 91}, 037903 (2003).

\bibitem{ra-prl-95-100501}
T.~C. Ralph, A.~J.~F. Hayes, A.~Gilchrist, {\it Phys. Rev. Lett.\/} {\bf 95},
  100501 (2005).

\bibitem{br-prl-95-010501}
D.~E. Browne, T.~Rudolph, {\it Phys. Rev. Lett.\/} {\bf 95}, 010501 (2005).

\bibitem{wa-nat-434-169}
P.~Walther, {\it et~al.\/}, {\it Nature\/} {\bf 434}, 169 (2005).

\bibitem{ki-prl-95-210502}
N.~Kiesel, {\it et~al.\/}, {\it Phys. Rev. Lett.\/} {\bf 95}, 210502 (2005).

\bibitem{pr-nat-445-65}
R.~Prevedel, {\it et~al.\/}, {\it Nature\/} {\bf 445}, 65 (2007).

\bibitem{lu-nphys-3-91}
C.-Y. Lu, {\it et~al.\/}, {\it Nature Physics\/} {\bf 3}, 91 (2007).

\bibitem{kn-nat-434-39}
E.~Knill, {\it Nature\/} {\bf 434}, 39 (2005).

\bibitem{ni-pra-71-042323}
M.~A. Nielsen, C.~M. Dawson, {\it Phys. Rev. A\/} {\bf 71}, 042323 (2005).

\bibitem{va-quantph-0702044}
M.~Varnava, D.~Browne, T.~Rudolph, {\it arXiv:quant-ph/0702044\/}  (2007).

\bibitem{da-pra-73-052306}
C.~M. Dawson, H.~L. Haselgrove, M.~A. Nielsen, {\it Phys. Rev. A\/} {\bf 73},
  052306 (2006).

\bibitem{scirevnote}
Quantum mechanics tells us that if a particular event can happen in two or more
  indistinguishable ways we must first sum the probability amplitudes before
  squaring to obtain a probability. Because these amplitudes can be negative,
  it is possible for events that we would intuitively expect to be possible to
  have zero probability. If the BS is more or less reflective than 50\% these
  two amplitudes no longer have the same magnitude, but are still opposite in
  sign, meaning that the probability is nonzero, but less than one would
  naively expect.

\bibitem{sps}
See special issue: Focus on Single Photons on Demand, Eds. P. Grangier, B.
  Sanders, and J. Vuckovic, New J. Phys. 6 (2004).

\bibitem{sa-nat-419-594}
C.~Santori, D.~Fattal, J.~Vu{\"{c}}kovi{\"{c}}, G.~S. Solomon, Y.~Yamamoto,
  {\it Nature\/} {\bf 419}, 594 (2002).

\bibitem{be-nat-440-779}
J.~Beugnon, {\it et~al.\/}, {\it Nature\/} {\bf 440}, 779 (2007).

\bibitem{ma-nphys-3-538}
P.~Maunz, {\it et~al.\/}, {\it Nature Phys.\/} {\bf 3}, 538 (2007).

\bibitem{spd}
See special issue: Single-photon: detectors, applications, and measurement
  methods, Eds. A. Migdal and J. Dowling, J. Mod. Opt. 51 (2004).

\bibitem{fr-pra-70-062302}
J.~D. Franson, B.~C. Jacobs, T.~B. Pittman, {\it Phys. Rev. A\/} {\bf 70},
  062302 (2004).

\bibitem{ba-pra-71-060302}
S.~D. Barrett, {\it et~al.\/}, {\it Phys. Rev. A\/} {\bf 71}, 060302 (2005).

\bibitem{li-pra-73-012304}
Y.~L. Lim, S.~D. Barrett, A.~Beige, P.~Kok, L.~C. Kwek, {\it Phys. Rev. A\/}
  {\bf 73}, 012304 (2006).

\bibitem{mbqc}
See forthcoming special issue: Measurement-Based Quantum Information
  Processing, Eds. J.-W. Pan and T. Rudolph, New J. Phys. (2007).

\bibitem{gi-pra-75-052328}
A.~Gilchrist, A.~J.~F. Hayes, T.~C. Ralph, {\it Phys. Rev. A\/} {\bf 75},
  052328 (2007).

\bibitem{ki-quantph-0611140}
K.~Kieling, T.~Rudolph, J.~Eisert, {\it arXiv:quant-ph/0611140\/}  (2006).

\bibitem I
I am indebted to all my collaborators past and present who have helped in my understanding of this subject. Optical quantum computing research at Bristol is
supported by the U.K. Engineering and Physical Sciences
Research Council, the U.K. Quantum Information
Processing Interdisciplinary Collaboration, the U.S.
Disruptive Technologies Office, the E.U. Integrated
Project Qubit Applications, the Leverhulme Trust, and the
Daiwa Anglo-Japanese Foundation.

\end{thebibliography}
\end{document}